\documentclass{iau}
\usepackage{graphicx}

\title[Comparison of Spiral Nuclear Regions] 
{Lessons  from  Comparisons between the Nuclear  Region  of   the  Milky  Way \&   
Those in Nearby  Spirals}

\author[J. S. Gallagher, T. M. Yoast-Hull \& E. G. Zweibel]   
{John S. Gallagher III$^1$, Tova M. Yoast-Hull$^2$
 \and Ellen G. Zweibel $^{1,2}$}

\affiliation{$^1$Department of Astronomy, University of Wisconsin Madison, WI 53706 USA\\ 
$^2$ Department of Physics and Center for Magnetic Self Organization in Laboratory and Astrophysical Plasmas, University of Wisconsin,  Madison, WI 53706 USA \\ 
email: {\tt jsg@astro.wisc.edu, yoasthull@wisc.edu, 
zweibel@astro.wisc.edu}}

\pubyear{20xx}
\volume{xxx}  
\pagerange{119--126}
\jname{IAU303}
\editors{TBD}
\begin{document}

\maketitle

\begin{abstract}
The Milky Way appears is a typical barred spiral, and comparisons can be made between its nuclear region and those of structurally similar nearby spirals.  Maffei 2, M83, IC~342 and NGC~253 are nearby systems whose nuclear region properties contrast with those of the Milky Way.  Stellar masses derived from NIR photometery, molecular gas masses and star formation rates allow us to assess the evolutionary states of this set of nuclear regions.   These data suggest similarities between nuclear regions in terms of their stellar content while highlighting significant differences in current star formation rates. In particular current star formation rates appear to cover a larger range than expected based on the molecular gas masses.  This behavior is consistent with nuclear region star formation experiencing episodic variations.  Under this hypothesis the Milky Way's nuclear region currently may be in a low star formation rate phase.
\end{abstract}

\firstsection 
\section{Introduction}

While the nuclear zone of the Milky Way can be studied in exquisite detail, it is a single example.  We therefore want to understand which of its features are unique to the Milky Way and which are generally found in nuclear regions of spiral galaxies.  The Milky Way's nuclear region, that we take to be the extent of  the central molecular zone (CMZ), can be characterized by a modest star formation rate (SFR), quiet super-massive black hole (SMBH) accreting at very low levels, and indications for a past nuclear outflow or wind in the form of the Fermi bubble (Su et al. 2010). Is this a normal mix of conditions for spiral CMZs, and if so how do they vary with the evolutionary phases of the regions?  Our group is particularly interested in how conditions in the CMZs of galaxies affect their high energy $\gamma$-ray emission and potential as sources of high energy neutrinos (see Yoast-Hull et al. this volume). 

Here we address these issues through comparisons between the Milky Way's CMZ and those of four nearby (D $\leq$ 4.5~Mpc) barred spiral galaxies. We focus on star forming properties, as primary drivers of the observable properties in these inactive nuclei. While most massive galaxies contain nuclear star clusters, probably inhabited by SMBH and containing some young stars , the level and extent of star formation in their CMZs varies. For example, while nuclear star clusters of spiral galaxies, including that in M31, frequently contain young stars e.g., \cite[Matthews et al. 1999]{matthews99},  \cite[Walcher et al. 2006]{walcher06}, and \cite[Lauer et al. 2012]{lauer12}, their CMZs cover a large range of conditions, including being all but non-existent.  The galaxies in our sample all have CMZs resembling that in the Milky Way in containing molecular gas and supporting star formation.

\section{Galaxy Sample}

Here we briefly summarize the characteristics of the galaxies and their CMZs.  Distances and total optical absolute magnitudes were taken from the literature. We measured H-band absolute magnitudes for each CMZ via aperture photometry on images in the 2MASS {\it Large Galaxy Atlas} (Jarrett et al. 2003).   The 2MASS images were also used to derive updated structural types (see Figure~1). Note that the CMZs are resolved and stand out in surface brightness in all 4 systems. The presence of bars is especially relevant since the Milky Way is a barred galaxy and bars have the ability to foster gas flows in nuclear regions.  

An assessment of the evolutionary rates of this sample rests on measuring SFRs and understanding the mass balance and structure of the ISM in the CMZ.  The former is difficult due to the practical issue of determining SFRs in dense and often highly obscured regions. The mass balance of the ISM is influenced by mass inflows (e.g., via bars) and outflows (e.g., from winds) from the CMZs, while structural data require mapping of complex, multi-phase ISM.  For this initial study we therefore focused on the current mass of molecular material that we take from published values.

\begin{figure}[h]
\vspace*{-0.4 cm}
\begin{center}
 \includegraphics[width=3.0in]{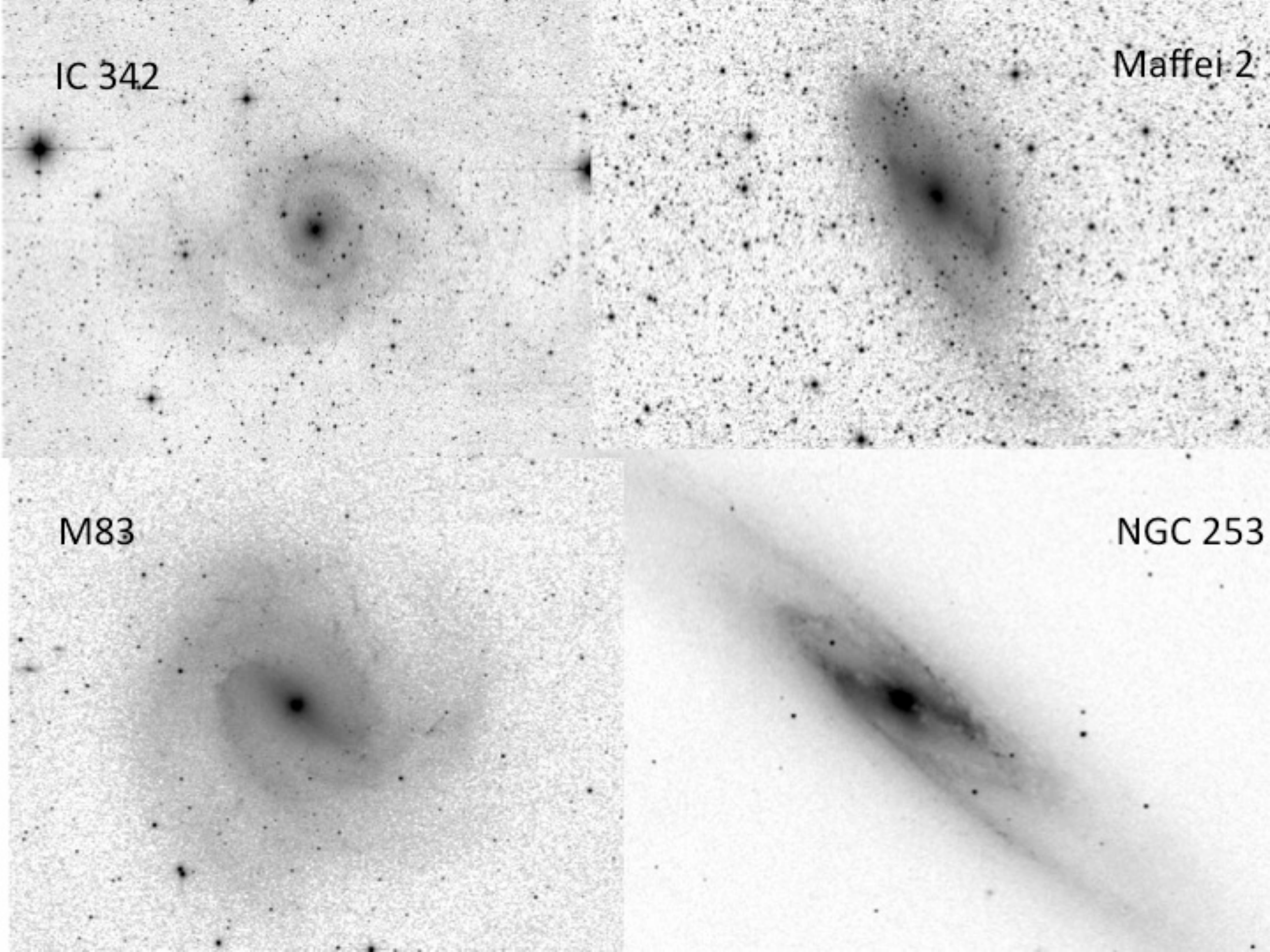} 
\vspace*{-0.2 cm}
 \caption{Sample of nearby barred spirals for comparison with the Milky Way. These H-band images are from 
the 2MASS survey (Jarrett et al. 2003).}
   \label{fig1}
\end{center}
\end{figure} 

{\it Milky Way:}  We adopt a classification of SABbc: for the structure of the Milky Way and assign it an absolute magnitude of M$_B \sim -$21.  Even though several details remain unclear, our home galaxy is a typical giant spiral.  We also adopt M$_{mol} \approx$3 $\times$10$^7$~M$_{\odot}$ for the Milky Way's CMZ. 

{\it IC~342:} This lumnous SABcd galaxy is weakly barred  at a distance of 3.5~Mpc and has an absolute magnitude of M$_B \approx -$22.  The CMZ M$_{mol} \approx$2 $\times$10$^7$~M$_{\odot}$ is from Schinnerer et al. (2003).

{\it Maffei~2:}  Although highly obscured by the Milky Way in the foreground, near-infrared images show Maffei~2 to be a well organized SBbc: spiral at D$=$3.9~Mpc. Due to the large foreground extinction, the optical magnitude of Maffei~2 is highly uncertain, but it probably is the least luminous galaxy in our sample. The  CMZ molecular disk is located within a gas-rich bar and contains M$_{mol} \approx$5$\times$10$^7$~M$_{\odot}$ of gas (Meier et al. 2008). 

{\it M83:} The most distant member of our sample at D$=$4.5~Mpc, M83 is an SBc system with M$_B \approx -$21. Its nucleus recently supported a starburst that produced a UV-bright region of young stars. The M83 CMZ has M$_{mol} \approx$5 $\times$ 10$^7$~M$_{\odot}$ (Muraoka et al. 2009). 

{\it NGC~253:} NGC~253 is a luminous, M$_B \approx -$22 SBc system with active star formation across the galaxy and especially in its starburst nucleus.  We adopt D$=$3.9~Mpc.  The galaxy is highly inclined and the starburst CMZ also shows an inclined, disky structure (Sakamoto et al. 2011). Estimates for M$_{mol}$ range from $\approx$10$^8$~M$_{\odot}$ from Sakamoto et al. to 3 $\times$ 10$^8$~M$_{\odot}$ derived by Weiss et al. (2008) from a submillimeter dust mass determination.  Sakamoto et al. also note the similarity between the molecular structures of the NGC~253 and Milky Way CMZs.

Star formation rate estimates are derived from thermal infrared and radio fluxes in the literature converted to SFRs following the prescriptions of Kennicutt \& Evans (2012), and are uncertain by factors of a few. For example, the CMZ SFR in M$_{\odot}$~yr$^{-1}$ for the Milky Way cover the range of 0.05$\leq$ SFR $\approx$ 0.15. For this pilot project we adopted estimates of the SFRs of 0.1 M$_{\odot}$~yr$^{-1}$ for IC~342 and Maffei~2, 0.5-1 M$_{\odot}$~yr$^{-1}$ for M83, and 3-10 M$_{\odot}$~yr$^{-1}$ for NGC~253.  Our future work on this topic will incorporate more systematic determinations of the CMZ SFRs. 

\section{Discussion}

We see that SFRs vary by factors of at least $\sim$100 while total molecular gas mass in the CMZ varies by only a factor of $\sim$10; the significant differences in the star formation efficiency are shown in Figure~2 that also illustrates the factor of 10 spread in SFRs at similar M$_{mol}$.  A simple Kennicutt-Schmidt near linear relationship between gas surface density and SFR evidently does not readily apply in these CMZs.  Figure~3 shows a comparison between the $\gamma$-ray flux for the Milky Way and NGC~253 obtained by assuming both systems are at the distance of NGC~253. The offset between the two galaxies is a factor of $\sim$10$^3$, comparable to the ratio of their SFRs but much larger than the factor $\sim$10 difference in gas mass (see also Yoast-Hull et al., these proceedings).

\begin{figure}[h]
\vspace*{-0.4 cm}
\begin{center}
 \includegraphics[width=2.75in]{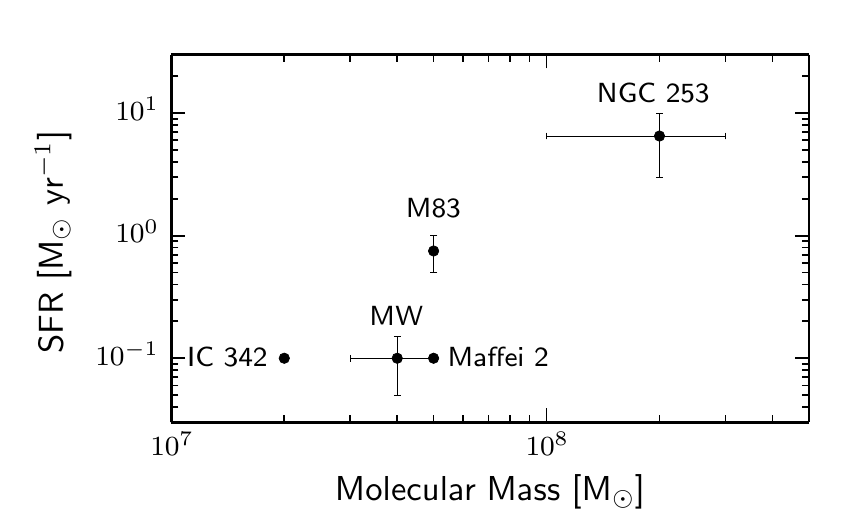} 
\vspace*{-0.2 cm}
 \caption{This plot shows the scatter in the estimated SFRs and molecular masses for CMZs of the galaxies in our sample. A few examples of error bars also are shown.}
   \label{fig2}
\end{center}
\end{figure}

Time scales to exhaust current gas supplies assuming that star formation is the only sink extend from $\sim$0.3-0.5~Gyr for the Milky Way and Maffei~2 to $<$100~Myr for M83 and NGC~253.  Outflows via winds will further reduce time scales to exhaust nuclear region gas supplies. The CMZ of NGC~253 supports a strong wind (e.g., Westmoquette et al. 2011), reducing the time scale to exhaust M$_{mol}$ in NGC~253 to $\sim$10~Myr, or roughly the age of the present starburst, in the absence of gas inflows.  The short gas exhaustion time scales indicate that gas flows into the centers of galaxies are required to fuel star formation. 

The H-band absolute magnitudes of the CMZs in this sample, (aside from the Milky Way where we lack this information, are -19.9$\pm$0.1 except for IC342 whose nucleus appears to be about 1 magnitude fainter.  Although interstellar obscuration and stellar population ages naturally complicate the relationship between absolute H-band magnitude and stellar mass, we simply assume that the similar H-band luminosities of the CMZs indicate similar stellar masses of $\sim$ 10$^9$~M$_{\odot}$.  In this case the lifetime average SFRs of the CMZs also must be similar at about 0.1~M$_{\odot}$~yr$^{-1}$ . This point and the short gas exhaustion time scales indicate that current SFRs do not necessarily reflect lifetime means for these CMZs.

This exploratory investigation thus suggests that star formation is episodic in the nuclei of galaxies like the Milky Way.  This possibility is not new (e.g., Loose, Kr\"{u}gel, \& Tutukov 1982) and also has support from recent studies (Schinnerer, B\"{o}ker, \& Meier 2003, Su, Slatyer, \& Finkbeiner 2010; also Bally et al. and Su et al. in this volume).  The CMZ of the Milky Way has the lowest star forming activity in the sample, and thus is likely to be in a low SFR evolutionary phase.  


\begin{figure}[h]
\vspace*{-0.4 cm}
\begin{center}
 \includegraphics[width=3.2in]{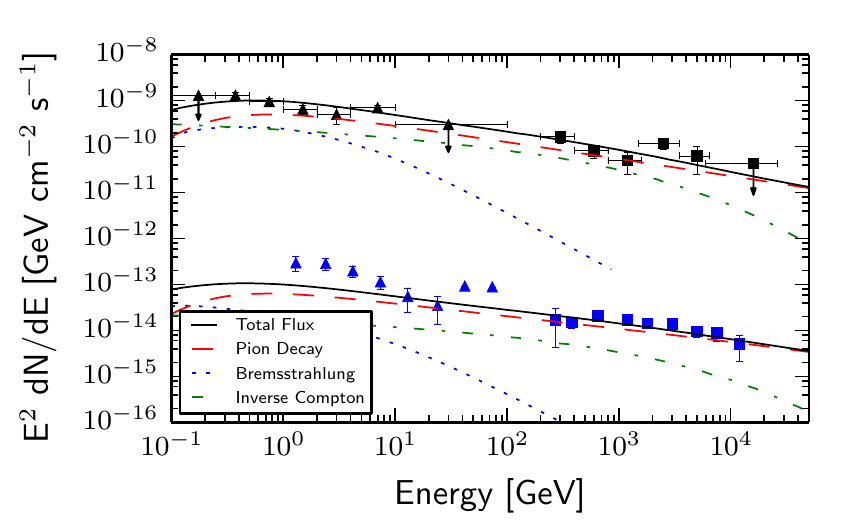} 
 \vspace*{-0.3 cm}
 \caption{Fluxes of $\gamma$-rays are shown for the CMZs of the Milky Way and NGC~253 assuming both galaxies are at the same distance.  The $\gamma$-ray data and models for their emission are described by Yoast-Hull et al. (2014) and Yoast-Hull et al. in this volume. }
   \label{fig3}
\end{center}
\end{figure} 

In the future it will be useful to develop consistent sets of measurements for the CMZs in nearby galaxies.  For example, deriving SFRs and M$_{mol}$ from similar types of observations will help to confirm the veracity of differences in the properties of nuclear regions.  Better mass determinations, both from model fits to gas kinematics and from stellar population spectral synthesis based on near infrared spectroscopy also will be valuable.  In addition the structures of the CMZs matter in building an improved understanding of how their various components interact, including cosmic rays and the central SMBH.  A better knowledge of the nuclear zones in neighboring galaxies, both with and without  AGN, will enhance the value of explorations of the central zone of the Milky Way.

{\bf Acknowledgements}  This work was supported in part by NSF AST-0907837, NSF PHY-0821899 (to CMSO), and NSF PHY-0969061 (to the IceCube Collaboration).

\end{document}